\documentclass[prb,showpacs,twocolumn]{revtex4}
\usepackage{graphicx}
\usepackage{graphics}
\usepackage{amssymb}

\usepackage{amsmath}
\usepackage{multirow} 
\usepackage[squaren, Gray, cdot]{SIunits} 
\usepackage{units} 

\begin{document}

\preprint{preprint submitted to Journal of Magnetism and Magnetic Materials -\\
November 10, 2009}

\title{Direct and inverse measurement  of  thin films magnetostriction. }

\author{J.-Ph. \textsc{Jay}}
\email{jay@univ-brest.fr}
\author{F. \textsc{Le Berre}} 
\author{S. P. \textsc{Pogossian}}
\author{M.V. \textsc{Indenbom}}


\affiliation{Laboratoire de Magn\'etisme de Bretagne - CNRS FRE 3117\\
Universit\'e de Brest, Universit\'e Europ\'eenne de Bretagne-
\\6, Avenue le Gorgeu C.S.93837 - 29238 Brest Cedex 3 - FRANCE.\\}



\date{\today}


\begin{abstract}
Two techniques of measurements of thin film magnetostriction are compared: direct, when changes of the substrate curvature caused by the film magnetization are controlled, and inverse ("indirect"), when the modification of the magnetic anisotropy induced by the substrate deformation (usually bending)  is measured. We demonstrate how both the elastic strength of the substrate and the effective magneto-mechanical coupling between the substrate deformation and magnetic anisotropy of the film  depend on different conditions of  bending.
Equations to be used for  magnetostriction value determination in typical cases are given and critical parameters for the corresponding approximations are identified.

\end{abstract}



\maketitle




\section{Introduction}
\label{}

There are two alternatives for magnetostriction measurements  of bulk ferromagnetic materials: it can be measured either directly, as it was initially observed from deformation under the magnetic field, or by an inverse way measuring   the magnetic anisotropy induced by mechanical stress applied to the sample \cite{Chikazumi}.
In thin films the magnetostriction deformations are hindered by the substrate and only a secondary  much smaller substrate deformation can be detected. By analogy to the bulk case, the direct technique uses the substrate deformation caused by the modification of the film magnetization. The inverse technique (often called "indirect", the term we find not very adequate) implies  studies of the thin film anisotropy changes caused by the deformation of the substrate (see \cite{Sander-RPP99} and references therein).

Description of these techniques involves  elastic properties of both the thin film and the substrate and, in particular for the direct technique, it took a long time to find the right equation to extract the magnetostriction parameters
\cite{Gontarz-64, Klokholm-76,lacheisserie-jmmm94,Trem-JMMM96,Marcus_JMMM_97}. The entire mechanical problem  is so complex that even after the right  solution was well established \cite{Marcus_JMMM_97} other contradicting theories are published \cite{wetherhold-jap03}. The theory for the inverse method appears to be simpler because the shape of the substrate is controlled directly and one can consider only the thin film properties. Nevertheless, different expressions of the stress induced anisotropy can be found in different papers without a clear explanation  of their origin \cite{Markham_IEEE_89, Ali-98,Cooke_JPD_00, Kundys-02}. The difficulties come from an interplay between longitudinal and transverse deformations in response to uniaxial stresses related by Poisson's ratio.
The problem is crucial for  applications of cantilevers  as sensors and actuators where the  role of the Poisson's ratio was specially investigated \cite{Iannotti-JAP05}.

The rigorous solutions for the bimorphous plate bending includes so heavy and not transparent equations  that for scientists of the magnetic community it is   difficult to understand the reason why the Poisson's ratios of the substrate and of the film appear in so different forms. At the same time the direct experimental verification of the theory is not convincing  due to the fact that the parameters of thin films can vary significantly compared to the bulk ones \cite{Trem-JMMM96}.

In this paper  without repeating the complete derivation we present simple criteria for choosing the correct equation in every practical case.
We pay particular attention to the limits of the validity of known solutions and we carry out comparative  measurements  of  magnetostriction of the \emph{same} thin film sample using both direct and inverse techniques.

\section{Revision of the theory}
\begin{figure}[h]
\centering
\includegraphics[angle=0,width=6.5cm]{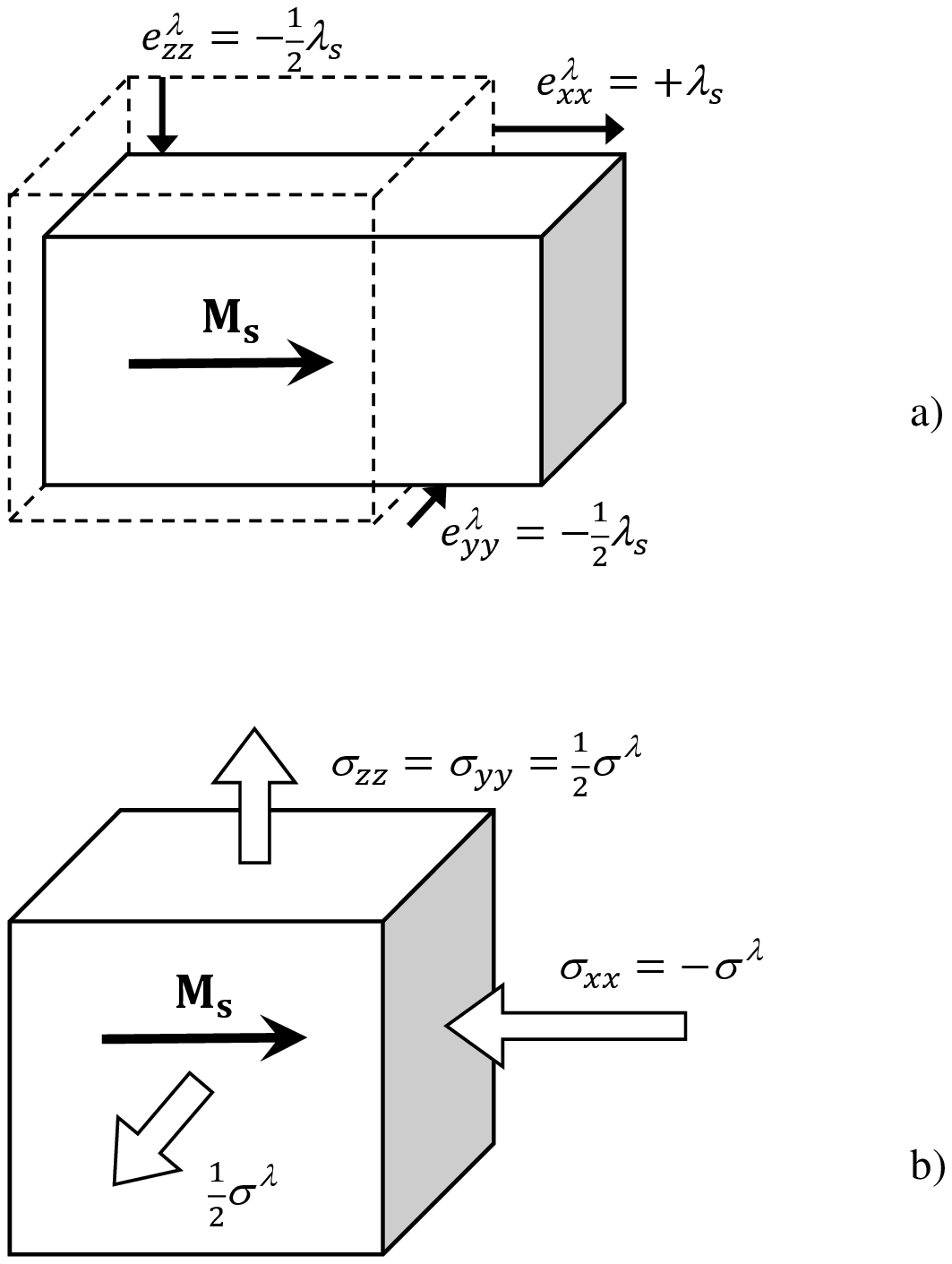}
\caption{ Comparison of two definitions of the anisotropic magnetostriction.
a) Spontaneous magnetostrictive deformations related to the magnetization vector oriented along $x$ axis. The relative deformation $e_{xx}^\lambda$ defined as the magnetostriction constant $\lambda_s$ is accompanied by half smaller opposite transverse deformations $e_{yy}^\lambda$ and $e_{zz}^\lambda$. The case of positive $\lambda_s$ when the initial cube of isotropic material is transformed to the elongated parallelepiped
is shown.
b) Alternatively the origin of magnetostriction can be defined trough the appearance of internal stresses $\sigma_{xx}^\lambda= \sigma^\lambda$, $\sigma_{yy}^\lambda= \sigma_{zz}^\lambda= -\nicefrac{1}{2}~\sigma^\lambda$ which can be equilibrated by external stresses (big white arrows) $\sigma_{ij}= -\sigma_{ij}^\lambda$ to block  the deformation of the material and, thus, to provide the elastic deformations opposite to the magnetostrictive ones.
}
\label{fig_MS1}
\end{figure}

Already the existence of two "alternative" theories of the magnetostrictive effect creates some confusions. The anisotropic magnetostriction can be introduced in two ways. First, as  the proper material deformations $e_{ij}^\lambda$ seen experimentally  when the sample reaches the  magnetic saturation (traditional technical definition \cite{Chikazumi}). In the simplest case of an isotropic material $e_{ij}^\lambda$ are related only to the magnetization direction ( $x$ axis in  Fig. \ref{fig_MS1}a) and are given by a single magnetostrictive coefficient $\lambda_s$:

\begin{eqnarray}
e_{xx}^\lambda&=& + \lambda_s \\ \nonumber
e_{yy}^\lambda&=& e_{zz}^\lambda=-\frac{1}{2} \lambda_s
\end{eqnarray}

In the second approach the magnetostriction is defined   as the proper  stresses originating from  internal magnetic  interactions  (first principle  approach \cite{lacheisserie-book}, see Fig. \ref{fig_MS1}b).For an isotropic material:
\begin{eqnarray}
\sigma_{xx}^\lambda&=& \sigma^\lambda \\ \nonumber
\sigma_{yy}^\lambda&=& \sigma_{zz}^\lambda= -\frac{1}{2}\sigma^\lambda
\end{eqnarray}

These definitions are equivalent providing that the  internal magnetostrictive stresses $\sigma_{ij}^\lambda$ produce elastic deformations equal to the magnetostrictive ones $e_{ij}^\lambda$, i.e.
\begin{eqnarray}
\sigma^\lambda= \lambda_s \frac{E}{1+\nu}
\label{eq-sigma-lambda}
\end{eqnarray}

where $\nu$ is the Poisson's ratio and $E$ is the Young modulus.

In general, the initial magnetic state of the sample is not known and the only way to obtain the magnetostriction value is to saturate magnetization successively in two orthogonal directions (say $x$ and $y$) and to measure the difference of deformation or stress  between these two states (i.e. $e_{xx}^\lambda - e_{yy}^\lambda = \nicefrac{3}{2}~\lambda_s$ or $\sigma_{xx}^\lambda~-~\sigma_{yy}^\lambda~=~\nicefrac{3}{2}~\sigma^\lambda$).

The corresponding   energy density of interaction between magnetostrictive deformation and elastic deformation (magneto-elastic energy) can be obtained from two equivalent general expressions :
$w_{ms}=- \sum_{ij} e_{ij}^\lambda \ \sigma_{ij} = -\sum_{ij} e_{ij}  \sigma_{ij}^\lambda $.

Very often the last expression is presented in a form where instead of $\sigma_{ii}^\lambda$ terms of type $B^{\gamma,2} \cos^2 \phi_i$ are used, where $\phi_i$ is the angle between the magnetization and the i-axis
 \cite{Kittel-RMP49,Trem-PRB95}.

Combining all together for our isotropic case we obtain  the angular variation of the magnetoelastic energy density for  magnetization rotation in the basal plane $(x,y)$:

\begin{eqnarray}
w_{ms}(\phi)
&=& - \frac{3}{2} \lambda_s (\sigma_{xx} -\sigma_{yy})\cos^2(\phi)  \nonumber \\
&=& - \frac{3}{2} \sigma^\lambda  (e_{xx}-e_{yy}) \cos^2(\phi)  \nonumber \\
&=& B^{\gamma,2} (e_{xx}- e_{yy}) \cos^2(\phi)
\end{eqnarray}
where $\phi$ is the angle between the magnetization and $x$ axis.

From this equation one can see that the physical meaning of $B^{\gamma,2}$ referred usually as the magnetoelastic coupling constant represents the characteristic magnetostrictive stress $\sigma^\lambda$ illustrated in figure \ref{fig_MS1}: $B^{\gamma,2} = - \nicefrac{3}{2} \ \sigma^\lambda = -\nicefrac{3}{2} \ \frac{E}{1+\nu} \lambda_s$.

For a thin magnetostrictive film deposited onto a substrate the planar components ($e_{xx}$,  $e_{yy}$ and $e_{xy}$) of the elastic deformation are fixed by the corresponding components of elastic deformations on the substrate surface.
For  a cantilever substrate  elastically bent along its length $x$   these elastic deformations are proportional to the curvature $1/R$.
A general expression of the energy of the unit area of such a sample in terms of  $1/R$ can be given as follows:\\
\begin{eqnarray}
W&=&W_{el}+W_{ms}+W_{an} \nonumber\\
 &=&\frac{1}{2} A d \left( \frac{1}{R}\right)^2 - B t \left( \frac{1}{R}\right)\cos^2 (\phi) - K t  \cos^2 (\phi)
\label{equ_nrj}
\end{eqnarray}

where $W_{el}$  is the integral  of the elastic energy density over the substrate thickness $d$; $W_{ms} = t\ w_{ms} $  ($t$ - thin film thickness) and the magnetic anisotropy energy density of the film is $W_{an}/t$.

For simplicity we consider only the uniaxial magnetic anisotropy   with easy  axis along the sample length. Here $K$ is the  usual anisotropy constant.

By analogy with the bulk measurements the direct way to get the films  magnetostriction  is to turn the magnetization direction from $\phi=0$  to 90$^\circ$ and to observe the corresponding change of the equilibrium substrate curvature that can be calculated by minimization of eq. (\ref{equ_nrj}):
\begin{eqnarray}
\Delta \left( \frac{1}{R} \right)= \frac{B \  t}{A \ d}
\label{equ_Delta}
\end{eqnarray}

 Practically this effect can be measured by deviation of a laser beam reflected from a point near to the free sample end:  $\Delta \theta =2 L\Delta \left( \frac{1}{R} \right)$. Here $L$ is the sample length (more precisely  the distance between the support and the laser spot on the sample).

Alternative way to measure the magnetostriction is the more frequently used inverse technique:  $ \left(\frac{1}{R}\right)$ is modified and  according to eq. (\ref{equ_nrj}) the corresponding changes of the effective uniaxial anisotropy constant is measured:
\begin{eqnarray}
K_{eff}=K + B \frac{1}{R}
\label{equ_Keff}
\end{eqnarray}

As follows from eq. (\ref{equ_Delta}) and (\ref{equ_Keff}) both the direct  and the inverse  techniques should give \emph{the same result}: the value of coefficient $B$ where the magnetostrictive parameters of the film are "hidden".
The equivalence of  these techniques is rarely discussed because  they are practically never  used together.

A closer look to the problem shows that the parameters $A$ and $B$ may be different for different experiments and thus comparison of data obtained from different techniques  is not so obvious. Moreover  not only direct and inverse experiments should be compared with care but also data obtained from the same technique should be treated differently depending on the geometry of the substrate bending in the given experiment.

In general, the Young moduli and Poisson's ratios  of the substrate  and of the film material are different and should be distinguished when used together.
Below we demonstrate how different algebraic forms of the Poisson's ratio of the cantilever substrate  $\nu_s$ enters to $A$ and $B$  in two characteristic  cases: anticlastic bending and cylindrical bending. And we  show in a simple way why the Poisson's ratio of film  $\nu_f$ and  Young moduli of both the substrate $E_s$ and the film $E_f$ appear always in the same form for all kind of deformations. One can easily understand the variety of these  forms of  $\nu$ ($\nu_s$ or $\nu_f$) by reviewing characteristic  cases of elementary deformations presented in Appendix  \ref{app_Poisson}.

Small bending of a cantilever in one direction inevitably induces  an opposite bending in perpendicular direction in order to satisfy the absence of transversal stresses at its free side edges \cite{LandauL-elast}. In this  case the principal surface deformation  is followed by the transversal deformation  as for an elastic element under uniaxial stress (see configuration 1 in Appendix \ref{app_Poisson}).
These two opposite bending produces an anticlastic shape of the cantilever that  can not develop simultaneously with increasing deformation. Due to the corresponding geometrical constrains the process becomes non linear and  a much more complicated deformation appears for large cantilever curvature. This effect  is usually not  considered  in the papers on thin film magnetostriction while  it can be non negligible as shown below.

The conditions when bending can be considered small and the corresponding equations are linear can be found only in specialized papers on the mechanics of thin plates. The relevant critical parameter for thin cantilever with large length $L$ to width $W$  ratio  is $\beta\sim W/\sqrt{R d}$.

The linear transversal   anticlastic deformation  across the sample width $ z=y^2 \nu_s /R $  is valid when this value at the sample edge $y=W/2$ is  less than about $0.1d$ for $\nu\approx 1/3$ i.e.  $\beta <1$ (see Appendix \ref{app_antic}).

At larger $\beta$ the anticlastic profile cannot progress anymore in the same way
 because an additional edge axial tension created by it can no longer be neglected.   The shape of the plate flattens and for $\beta \gtrsim 3$ one can consider the bending as purely cylindrical. Another case of the  cylindrical bending   is when transverse deformation is blocked near to the sample fixed end  by the support (clamping effect \cite{Iannotti-JMMM99}). An immediate consequence of the fixed transversal deformations is the appearance of the transversal stresses $\sigma_{yy}$ (configuration 2 in  Appendix \ref{app_Poisson}).

It is obvious that both the effective elasticity coefficient  of the plate bending $A$ and the  coupling coefficient  $B$ between the bending and the magnetostrictive stress in the film are different in these two limit cases.\\

Pure bending in the principal cantilever direction $x$ produces the longitudinal  elastic deformations $e_{xx}(z)=z/R$ (see Fig. \ref{fig-jig}).

In the anticlastic case the elastic energy of the unit area of the curved plate is  \begin{eqnarray}
W_{el}=\int_{-d/2}^{d/2} \frac{1}{2} \sigma_{xx} \ e_{xx} \ dz = \frac{1}{24} E_s d^3 \left( \frac{1}{R}\right)^2
 \label{equ_Wel}
\end{eqnarray}

taking into account the relation between  $e_{xx}$  and   $\sigma_{xx}$   (configuration 1 in  Appendix \ref{app_Poisson}: $\sigma_{xx}=E e_{xx}$).
This gives  ${A=\nicefrac{1}{12}~ E_s d^2}$.

The magnetoelastic coupling energy of the thin film with planar magnetization is  fully determined by the  surface deformation of the much thicker substrate:

\begin{eqnarray}
  W_{ms}&=&t B^{\gamma,2} \left( e_{xx}- e_{yy} \right)_{z=d/2} \cos^2 \phi \nonumber \\
&=& t B^{\gamma,2} \frac{d}{2R}  \left(1+\nu_s \right) \cos^2 \phi
  \label{equ_Wms}
\end{eqnarray}
that gives $B=\nicefrac{1}{2}  \ d B^{\gamma,2}  \left(1+\nu_s \right)$.
Thus expression  (\ref{equ_Keff}) of the effective magnetic anisotropy for inverse magnetostriction measurements in the anticlastic case includes the Poisson's ratio of the substrate in the form $(1+\nu_s)$ .

From eq. (\ref{equ_Delta}) we obtain the variation of the curvature $1/R$ produced by  the magnetization rotation from $\phi=0$ to 90$^\circ$ to be used in the direct technique:

\begin{eqnarray}
\Delta \left( \frac{1}{R} \right)_{anticlastic}=\frac{B t}{A d} = 6 \ \frac{1+\nu_s}{E_s} \ \frac{t}{d^2} \ B^{\gamma,2}
\label{equ_Delta_antic}
\end{eqnarray}

This result exactly corresponds to the solution given by Tr\'{e}molet de Lacheisserie  \cite{lacheisserie-jmmm94}.

We should remember here that the Young's modulus $E_s$  and Poisson's ratio $\nu_s$  of the \emph{substrate} enters into $W_{el}$   and $W_{ms}$  while  $B^{\gamma,2}$ includes the Young's modulus $E_f$  and  the Poisson's ratio $\nu_f$  of the film.\\


For the cylindrical curvature, one have to use the relation between $\sigma_{xx}$ and $e_{xx}$ corresponding to the fixed transversal deformation ($e_{yy}=0$, configuration 2 in Appendix \ref{app_Poisson}: $\sigma_{xx}{=E/(1-\nu^2)~e_{xx}}$). In this case, the  integration  (\ref{equ_Wel}) gives harder plate bending coefficient $A= \nicefrac{1}{12}~ E d^2/(1-\nu^2)$. At the same time the elastic coupling becomes smaller $B=\nicefrac{1}{2} ~ d B^{\gamma,2}$. Consequently,  $\nu_s$ does not enter into the equation (\ref{equ_Keff}) of $K_{eff}$ used for inverse measurements unlike  the previous anticlastic case.

As to the direct measurements, eq. (\ref{equ_Delta}) provides:
\begin{eqnarray}
\Delta \left( \frac{1}{R} \right)_{cylindrical}=\frac{B t}{A d} = 6 \frac{1-\nu_s^2}{E_s } \frac{t}{d^2} B^{\gamma,2}
\label{equ_Delta_cyl}
\end{eqnarray}

This equation giving $(1-\nu_s)$ smaller substrate curvature than in the anticlastic case can be found in the literature,  often without any justification. In reality, it has no practical application  because  in direct measurements the cantilever curvature  is extremely small ($\beta \ll 1$) and clamping effect can be neglected for cantilevers with large $L/W$ ratio.

One should note that the elastic deformation $e_{xx}$ due to the plate bending is not the only deformation produced by the magnetostrictive stress during direct measurements of magnetostriction. This stress also changes the  sample  length creating homogeneous deformation $e_{xx}^L$. Consequently   the  position of the neutral line -- where total longitudinal deformation $e_{xx}(z) + e_{xx}^L=0$ -- is shifted as discussed by many authors (see for example \cite{lacheisserie-jmmm94,Marcus_JMMM_97,Iannotti-JMMM99}. The elastic energy corresponding to this additional deformation does not interfere with the energy of the pure substrate bending and, thus eq. (\ref{equ_Delta}) is unaffected.
The shift of the neutral line can be illustrated by considering separately symmetrical and antisymmetrical deformations produced by films deposited on both surfaces of substrates (see Appendix \ref{app_shape}). This small $e_{xx}^L \approx \lambda_s t/d$ usually does not produce any measurable effect.

\section{Experimental details} \label{section-exp}

We apply both direct and inverse techniques to the \emph{same} sample. For illustration we have chosen a \{[Tb$_{34\%}$Co$_{66\%}$]60 \AA/[Co$_{42\%}$Fe$_{58\%}$]50\AA\}$\times$ 10 multilayer deposited onto  rectangular Corning glass substrate ($22 \times 5 \times \unit{0.145}{\ \milli \cubic\meter} $) from CoFe and TbCo
mosaic 4 inch targets using a Z550 Leybold RF sputtering equipment
with a rotary table technique. An uniaxial anisotropy along sample length was induced by the field of a permanent magnet applied during deposition process.
The direct measurements were realized using cantilever deflectometry technique. Details of the sample preparation and of the cantilever measurements can be found in \cite{Jay-06}.

\begin{figure}[h]
\centering
\includegraphics[angle=0,width=7.5cm]{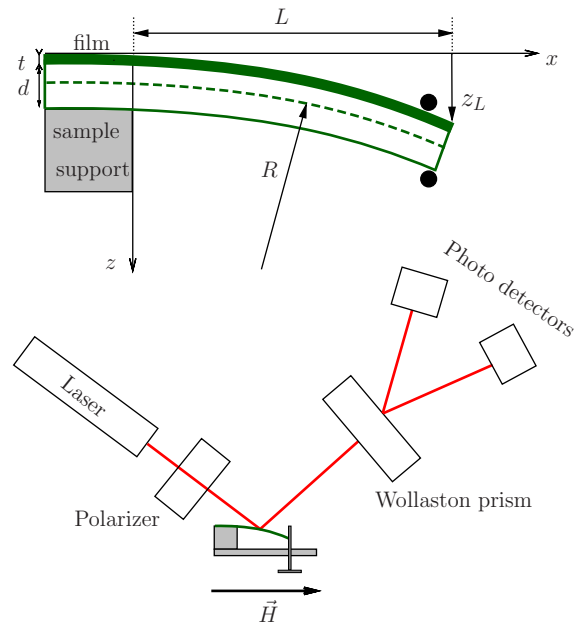}
\caption{(Color on line) Schematic diagram of the indirect measurements of the film magnetostriction using MOKE magnetometer. The geometry of sample bending is shown at the top of the figure. The dashed line shows the neutral line where the longitudinal deformation is zero.
 }
\label{fig-jig}
\end{figure}

For the inverse measurement we have studied local magnetization curves of bent sample using the magneto-optical Kerr effect (MOKE) as shown in figure \ref{fig-jig}.
We have chosen the longitudinal Kerr effect geometry
for which the planar magnetization component parallel to the plane of incidence is measured. The incident laser light is polarized vertically and  the light reflected from the film surface is split by a Wollaston prism to 2 photodetectors so that their signal difference is proportional to the Kerr rotation.
The incidence plane and the sample surface are parallel to the direction of
the  applied magnetic field. The sample can be rotated in its plane in order to measure the film local magnetization either along  its length (as shown in the figure) or along  its width.

The sample  is  mounted on a special  bending jig shown
Fig.\ref{fig-jig}. One end of the sample is glued on a support so that the film is on the top free surface from which the laser is reflected. The other sample end can be moved vertically ($z_L$) by a screw in order to bend it  up or down.
To respect the sign given in the theoretical equations we define positive curvature when the sample is bent downwards, and negative when the sample is bent
upwards.
We have found that the most precise way to get the sample curvature in our experimental setup is to analyze the modification of the reflected laser beam.
The light spot observed on a screen  after reflection from curved surface  is spread in two directions  proportionally to the principal curvatures.
This method is similar to the industrial multi-beam technique of measuring curved surfaces.
With a single beam we can measure the curvature at the same small area where the MOKE  hysteresis is studied.
In order to justify and calibrate our single beam technique we have analyzed the reflection from a number of cylindrical objects of known radius.
Measuring our sample we could see that the light reflected near to the support forms on the screen a narrow line thus showing that the sample  is curved here only along its length and remains flat in the transverse direction (cylindrical curvature).  Further from the support we could see that the laser spot becomes elliptical thus demonstrating the anticlastic shape of the sample deformation (see Appendix \ref{app_antic}).



\section{Results and discussion}

\begin{figure}[h]
\includegraphics[angle=0,width=8.5cm]{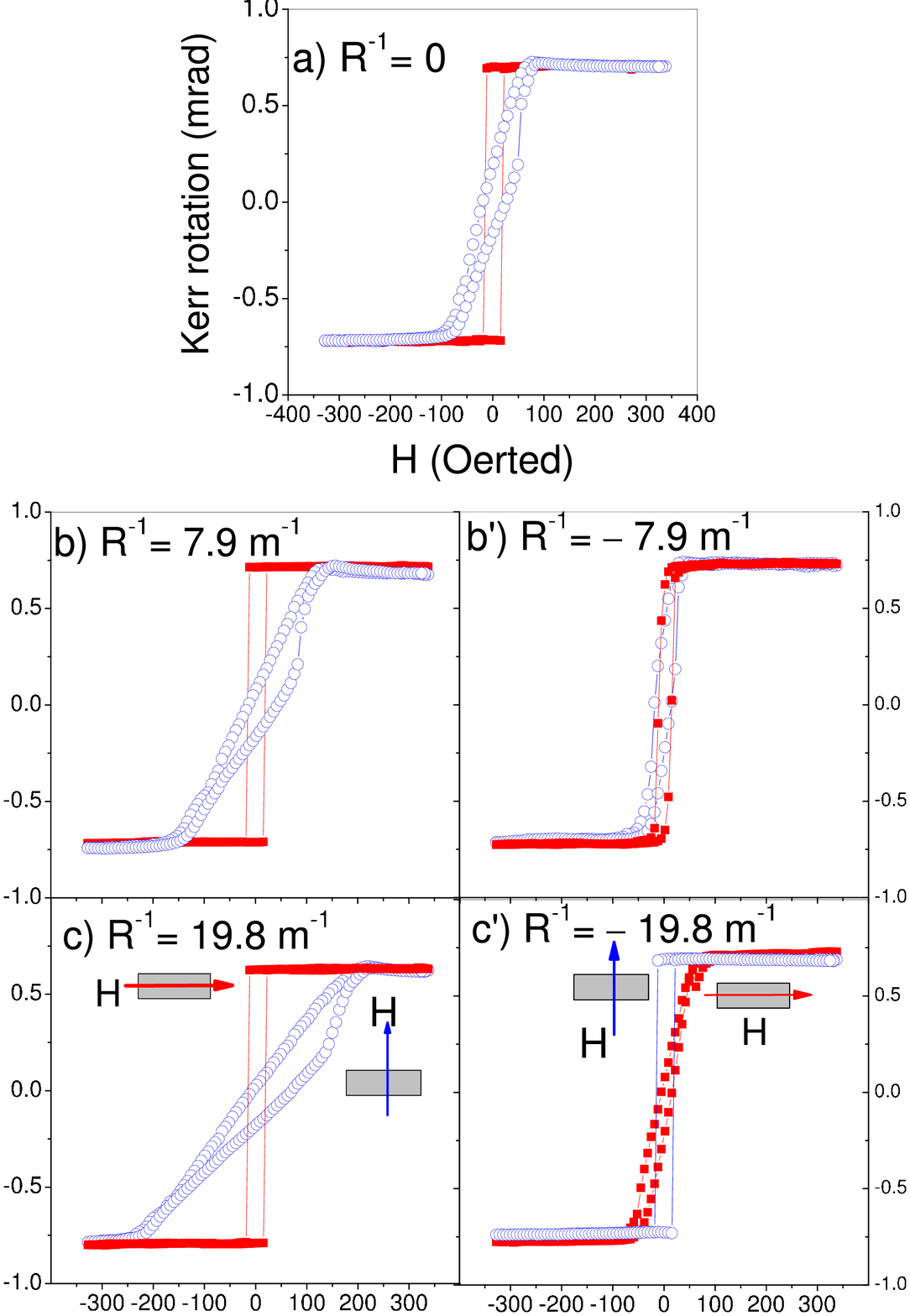}
\caption{ (Color on line)  MOKE magnetization loops measured close to the sample support ($x=\unit{2} \ \milli \meter$)  for different positive (left column) and negative (right column) cantilever curvatures.
 Red full square symbols - magnetic field applied along the sample. Blue open circle symbols - magnetic field applied across the sample
 as shown in c) and c') near to the corresponding curves}
\label{fig:moke2mm}
\end{figure}

We have studied the variation of the MOKE hysteresis with curvature at $x=\unit{2}{\ \milli \meter}$ from the support where the maximum curvature is achieved. Characteristic results from a sequence of  measurements  done for different bending  in positive and negative directions are shown in Fig. \ref{fig:moke2mm}.
For every curvature the sample has been rotated so as to obtain magnetization cycles  both for the  magnetic field  applied along the sample large axis  (full red square symbol) and perpendicularly to it (open blue circle symbol).

 As  observed with VSM similarly to our previous study \cite{Jay-06} the unstressed sample has a uniaxial anisotropy:  rectangular magnetization along the sample length  shows the easy axis and slanted magnetization along the sample width (hard axis magnetization) provides the initial anisotropy value (see Fig. \ref{fig:moke2mm}(a))

 For positive curvatures (left column), the large axis hysteresis loop shape does not vary, keeping the rectangular shape characteristic for the easy axis magnetization,  whereas the short axis hysteresis loop is getting more and more
slanted when curvature increases.  Thus we clearly see that a positive
curvature (stress) reinforce the anisotropy.
Negative curvature produces the opposite effect: the uniaxial anisotropy, first, is reduced by the applied stress (the slope of the short  axis hysteresis loop is reduced by stress counterbalancing the initial anisotropy) and, then, the easy anisotropy axis switches from longitudinal to transversal direction and the anisotropy increases again.
In our sample we observe  that this switching is not abrupt:
the long axis hysteresis starts to slant before the short axis hysteresis becomes completely rectangular (see  Fig. \ref{fig:moke2mm} (b')).  As we show below, this effect appears if the sample is not perfectly homogeneous.

\begin{figure}[h]
\includegraphics[angle=270,width=8.5cm]{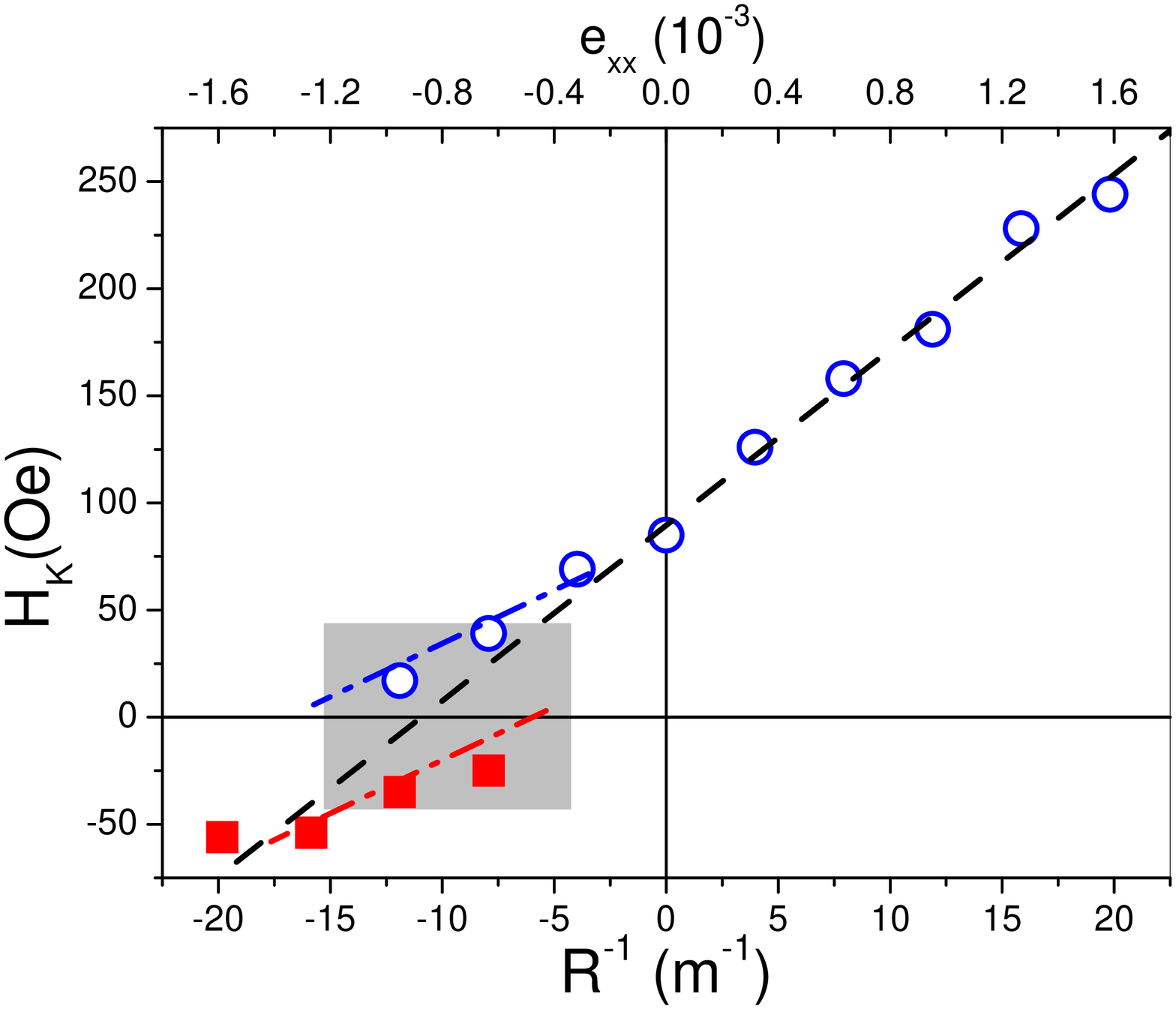}
\caption{(Color on line)  Dependence of anisotropy field measured as a function of curvature for fixed position ($x=\unit{2}{\ \milli \meter}$). Positive values of $H_K$ (open blue circles are obtained for $H \parallel x$) and negative values of $H_K$ (full red squares are obtained for $H \parallel y$).
 }
\label{fig-hk}
\end{figure}


Contrary to what was observed on the VSM cycles, back and forth branches in the hard axis loop are not fully superimposed and the cycle is somewhat asymmetrical.
This is not an unusual feature of MOKE measurements caused by  a non linear interference  with the  transverse magnetization signal \cite{Prokop-04}.
We have found that for  magnetostrictive films this MOKE loop feature
may have an additional reason. We observed a superposition of the magnetization and magnetostrictive response when the free sample end did not touch the screw at all and, thus, could move freely under the magnetostrictive stress in the film. Probably a small part of the latter effect can persist even when the sample end is touching the screw.

This  small distortion of the hard axis magnetization loop can be excluded by the following procedure of the determination of the uniaxial anisotropy field $H_K$. The loop is firstly "symmetrized" by
addition of the back and forth branches. Then the central part of
it $[-20$~Oe;~$+~20~$Oe$]$ is linearly fitted and $H_K$ is determined as the intersection of the fitted straight line with the  saturation magnetization level.
The  anisotropy field $H_K$ and the corresponding  anisotropy constant $K$ ($H_K= 2 K/(\mu_0 M_s )$ where $M_s$ is the saturation magnetization)  are defined positive when the hard axis loop is seen for applied field parallel to the small sample axis (see Fig. \ref{fig:moke2mm} a)-c), and negative if the hard axis loop is seen for applied field parallel to the large sample axis (see Fig. \ref{fig:moke2mm} c').


For a perfectly homogeneous  sample  with a well defined uniaxial intrinsic anisotropy direction (i.e. long sample axis), we would observe an abrupt reversal between slanted and rectangular loops at the moment of 90$^\circ$ rotation of the easy axis when the initial anisotropy is compensated by the stress induced anisotropy.
In reality, as we indicated above (Fig. \ref{fig:moke2mm}b') the transition is gradual. There is an interval around the compensation point where $H_K$ can be formally measured for both directions.
This is due to  small variations of both magnitude and direction of the intrinsic anisotropy in the sample area probed  by laser. This kind of anisotropy inhomogeneity was already observed and quantified in our previous work \cite{Jay-06}. There we have shown that large domains of opposite magnetization exist where maximum easy axis distribution is about $\pm$~6~$^\circ$ from the average direction over the whole sample.
In the intermediate region every  measured $H_K$ corresponds not only to the average anisotropy of given sign but also to its  relative  occurrence.

Figure \ref{fig-hk}  shows the evolution of the obtained positive and negative $H_K$ with the substrate curvature ($R^{-1})$ or the equivalent surface  strain $e_{xx}=d/2R$).

In order to get the  expected linear  variation of $H_k$  with curvature one has to consider only points outside the intermediate compensation region shaded gray in the figure.
Inside this region the slope is about twice smaller because the repartition between the areas with positive and negative $H_K$ also varies with curvature.
So if this region is not identified the resulting magnetostriction value can be considerably reduced.
The slope of the line fitting the points outside the compensation region $\frac{\partial H_K}{\partial (1/R)}=8.2$ Oe.m $=\unit{650} \ \ampere$ allows us to determine the magnetoelastic coupling coefficient of the film $B^{\gamma,2}$.

As  mentioned above for this measurement   the deformation is cylindrical  close to the support. So, for eq. (\ref{equ_Keff}) we have to take the corresponding form of the effective magnetoelastic coefficient  $B$:

\begin{eqnarray}
B^{\gamma,2}= \frac{\mu_0 M_s}{d} \frac{\partial H_K}{\partial (1/R)}
\end{eqnarray}
With $\mu_0 M_s= \unit{1.4}{\ \tesla}$    obtained from the magnetic measurements and $d=\unit{145}{\ \micro \meter} $
we found $B^{\gamma,2}=\unit{6.3}{\ \mega \pascal} $ .

Far from the support the  deformation of the sample  gets an anticlastic shape if the bending is not too strong (see Appendix \ref{app_antic}).
As was mentioned above,  we observe this anticlastic deformation as  the dispersion of the reflected laser beam: the laser spot is spread to an ellipse after reflection from the sample.
For the deformation  being equivalent to the deformation of narrow cantilevers the magnetoelastic coefficient $B$ is by $(1+\nu_s)$ larger. Correspondingly one have to use for this case another expression relating $B^{\gamma,2}$ and the induced anisotropy:

\begin{eqnarray}
B^{\gamma,2}= \frac{1}{1+\nu_s} \frac{\mu_0 M_s}{d} \frac{\partial H_K}{\partial (1/R)}
\end{eqnarray}


In order to observe the effect of this shape modification we have carried out measurements at different points from the support.

\begin{figure}[h]
\includegraphics[angle=0,width=8.5cm]{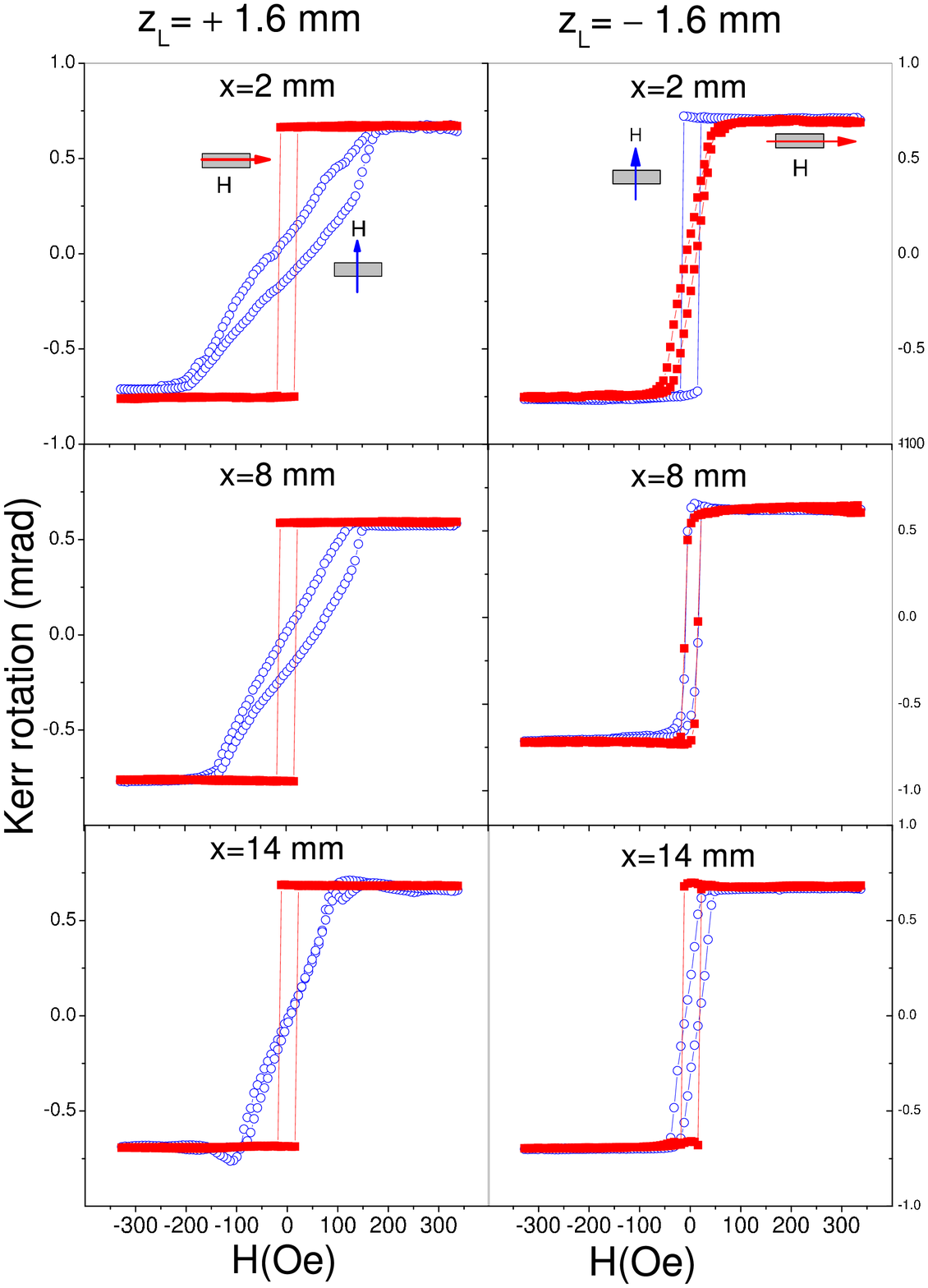}
\caption{ (Color on line)  MOKE magnetization loops measured for opposite sample bending $z_L=+\unit{1.6}{\ \milli \meter}$  (left column) and $z_L=-\unit{1.6}{\ \milli \meter}$  (right column) at different position on the sample: $x = \unit{2}{\ \milli \meter}$, $x = \unit{8}{\ \milli \meter}$  and $x = \unit{14}{\ \milli \meter}$.  Red full square symbols - magnetic field applied along the sample length and blue open circle symbols - magnetic field applied across the sample}
\label{fig-pm2-x}
\end{figure}

MOKE hysteresis loops obtained for a given deformation $z_L=+\unit{1.6}{\ \milli \meter}$, left column, and $z_L=-\unit{1.6}{\ \milli \meter}$ right column, when laser spot is swept along the sample length from a position close to the clamped tip ($x = \unit{2}{\ \milli \meter}$) to $x= \unit{14}{\ \milli \meter}$   trough $x= \unit{8}{\ \milli \meter}$ are presented Fig.  \ref{fig-pm2-x}.

For positive deviation (${z_L=+\unit{1.6}{\ \milli \meter}}$), the closer to the clamped sample tip (${x= \unit{2}{\ \milli \meter}}$) the more anisotropic the sample is:
 The small axis hysteresis loop is more slanted when recorded closer to the clamped tip. The hard axis (small side of the sample) is thus getting harder when $x$ decreases. $H_K$ as defined above remains positive decreasing with $x$  from its maximal value   to its initial value close to the free end.
The reason for this is that the created curvature and the corresponding induced anisotropy reduce progressively with the distance $x$ from  the support completely disappearing at the free end $x=L$ (see Appendix \ref{app_shape}).

\begin{figure}[h]
\includegraphics[angle=270,width=8.5cm]{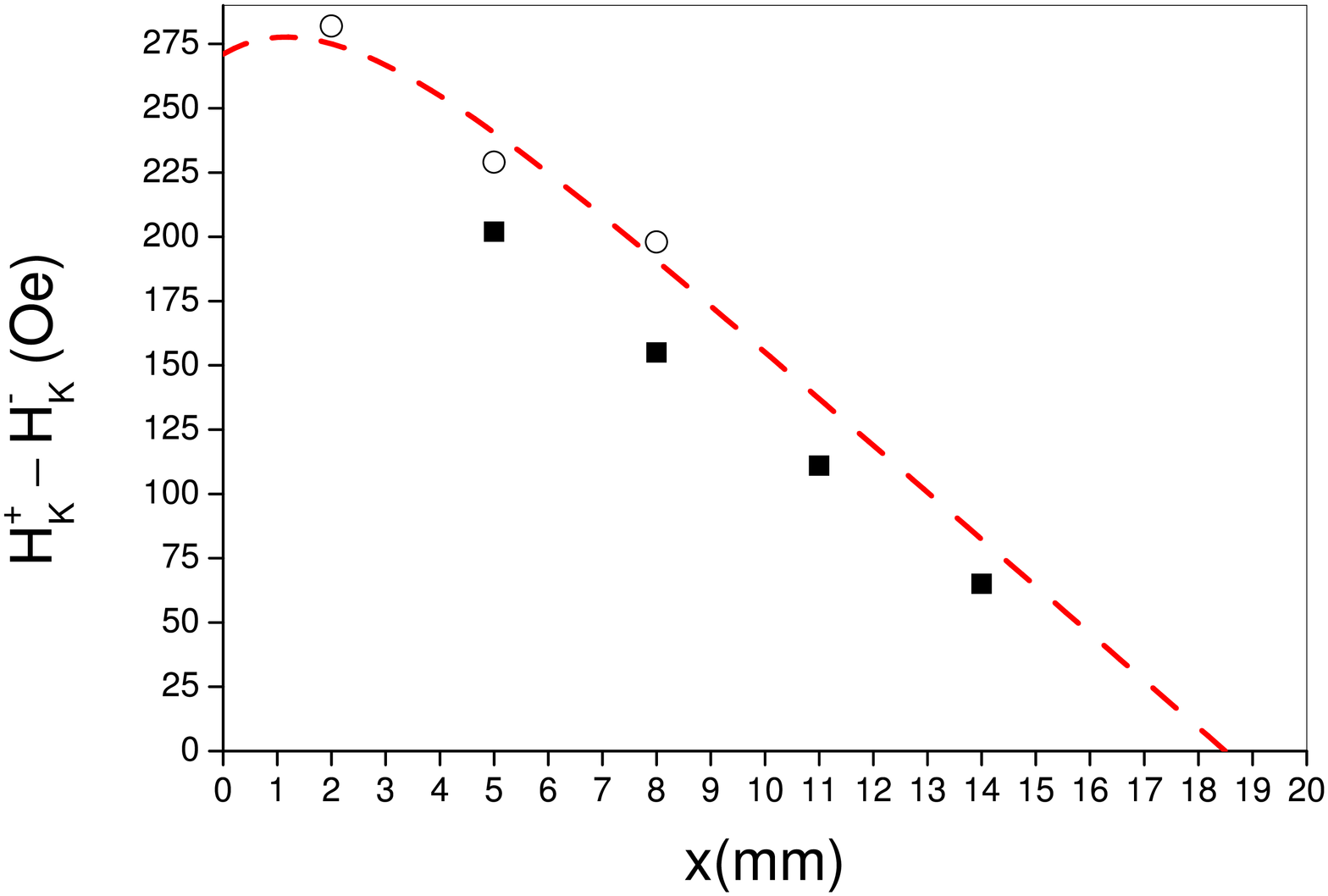}
\caption{(Color on line)
Modification of the induced anisotropy along the sample length. Open circles shows results involving anisotropy field $H_K^+ >0$  created by  positive bending and $H_K^- <0$ created by negative bending.  Results shown by full squares involve  positive anisotropy fields $H_K^+$  and $H_K^-$ for both opposite bending.
A theoretical  estimation taking into account the clamping effect is shown by the dashed line.
}
\label{fig-HK(x)}
\end{figure}

 For negative deviation ($z_L=- \unit{1.6}{\ \milli \meter}$) we observe the inversion of easy and hard direction at small $x$ where the induced anisotropy overcomes the initial one. Correspondingly, $H_K$ varies from a negative to a positive  value along the sample length. This transition is not abrupt similarly to the modification of $H_K$ as function of the bending curvature discussed above and there is an intermediate compensation area approximatively in the middle of the sample.
We represent the evolution of the induced anisotropy along the sample length by the difference between $H_K$ obtained for positive ($H_K^+$) and negative ($H_K^-$) bending (see Fig. \ref{fig-HK(x)}). Such presentation allows us to exclude the  inhomogeneity  of the initial  anisotropy of the undeformed sample.
We can distinguish two kinds of points: when $H_K^+$ and $H_K^-$ have different signs
in the part of the sample close to  the support (open circles in the figure) and when  $H_K^+$ and $H_K^-$ remain both positive in the second part of the sample (closed square in the figure). Similarly to the evolution with $1/R$  discussed before these two kinds of points coexist in the intermediate compensation area and expected theoretical value should lie somewhere between them.
Except of the inconvenience caused by  the sample inhomogeneity, there are two  effects  to be considered for quantitative analysis of these data:  the effect of the support clamping and the edge blocking for large curvature.

The cylindrical shape imposed by the clamping at the support extends to some distance  thus producing in this region the  strengthening of its bending and the reduction of the induced anisotropy as discussed above (see discussion before eq. (\ref{equ_Delta_cyl}). The total shape of the clamped sample does not depend on the bending amplitude until the curvature is small enough and the linear equations can be used.
The  variation of the induced anisotropy  due to clamping effect can be illustrated by eq. (\ref{equ_Knorm})  of Appendix \ref{app_shape}.

An example of such an estimation is shown by the dashed curve in Fig. \ref{fig-HK(x)}  calculated with the following parameters of our sample:
sample length $L=\unit{18.5}{\ \milli \meter}$,
$\nu_s=0.2$, characteristic length of the clamping effect $a \approx W/2 =\unit{2.5}{\ \milli \meter}$ and amplitude coefficient $6FB^{\gamma,2}/E_sW d^2=\unit{1.2 }{\ \times \megad \ampere \per \meter \squared}$. The theoretical predicted deviation from the linear dependence is clearly visible close to the sample support.

The maximum curvature,  for which the condition of this linear approximation   $\beta \sim W/\sqrt{R d} \lesssim 1$ is valid, for our sample is $(1/R)_{max} =6$ m$^{-1}$. In our experiment the largest  sample curvature considerably  overcomes this value near to the support (see Fig. \ref{fig:moke2mm}) and a further  gradual modification of the anisotropy with curvature should be added according to the calculated correction
$\kappa(1/R)$ (see Appendix \ref{app_antic}, eq. (\ref{equ-kappa}) and Fig. \ref{fig-z-kappa}b)). Since  $1/R$ reduces almost linearly along the sample length, the low curvature  approximation can be  used till maximal allowed bending only  close to the sample free end ($x \gtrsim \unit{14} \ \milli \meter$) where, unfortunately,  the induced anisotropy is too small for precise measurements of the magnetostriction.

Thus, in the main part of the sample we have a complicated superposition of at least three effects: the compensation effect with the revealed  inhomogeneous initial anisotropy, the clamping effect and the non linear shape effect.

Interestingly, our experimental measurements near to the cantilever support discussed above (Fig. \ref{fig:moke2mm}, \ref{fig-hk}) have no such complication. The clamping of the cantilever end imposes the cylindrical curvature at any width to length ratio and at any bending amplitude, and excludes all other effects considered above. Consequently the determination of  the magnetoelastic coefficient $B^{\gamma,2}$ has less uncertainty.

For our direct measurements of magnetostriction by deflectometry the anticlastic shape of the sample can be always considered: The aspect ratio $L/W\sim 4$ of the cantilever is sufficiently large to neglect the effect of clamping \cite{Iannotti-JMMM99} and the curvature is extremely small. Thus corresponding anticlastic parameter $\beta \lesssim  0.01$ and only eq. (\ref{equ_Delta_antic}) have to be used.

\begin{figure}[h]
\includegraphics[angle=270,width=8.5cm]{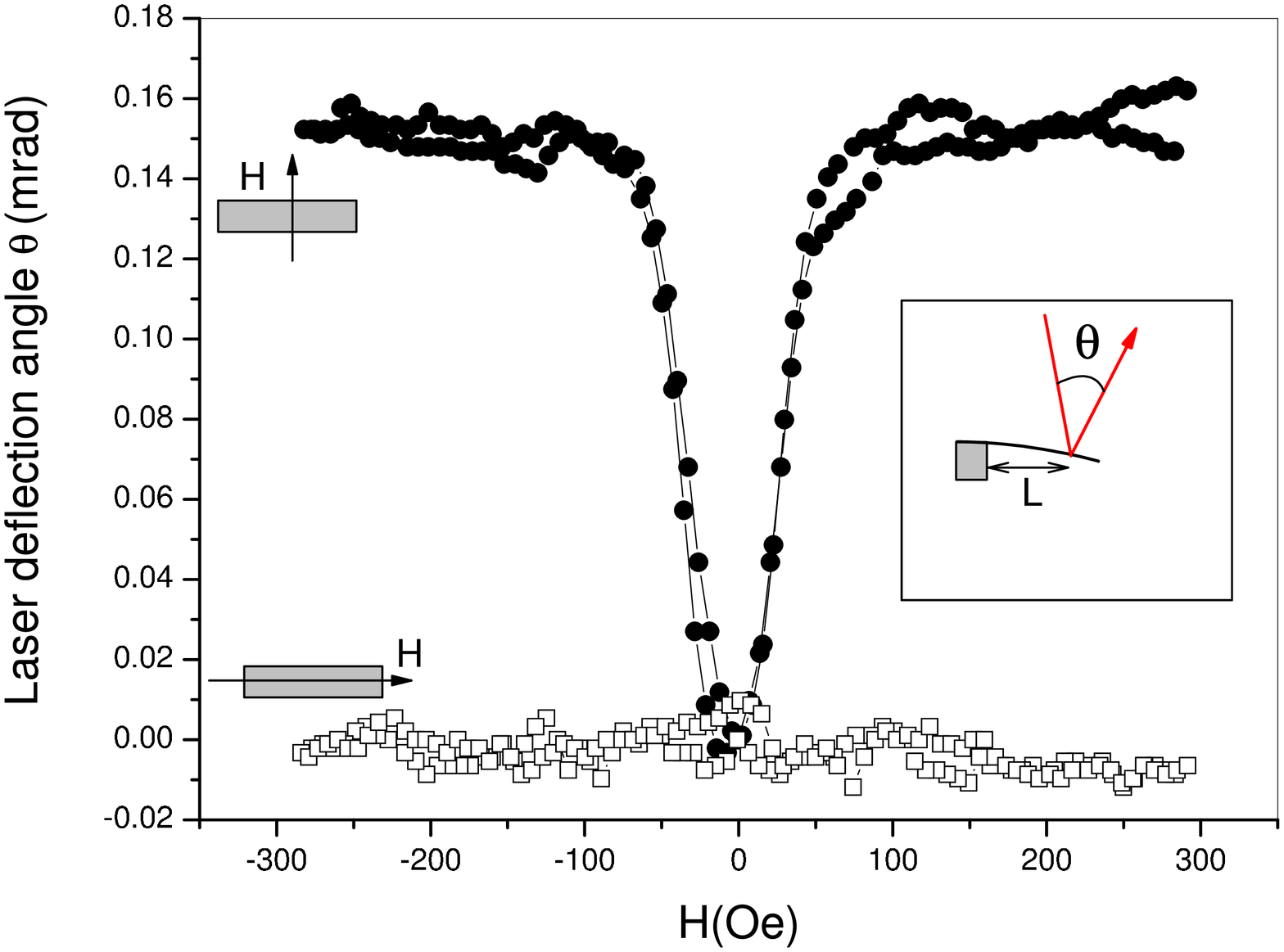}
\caption{ (Color on line)
Deflectometry measurements of the  magnetostrictive bending under the  field applied along the sample length and perpendicularly to it as schematically shown near  each corresponding curve. The geometry of the laser beam deflection is sketched in the inset.
}
\label{fig-defl}
\end{figure}

Figure \ref{fig-defl} shows deflectometry measurements of our sample  under the  magnetic field applied  in two directions: parallel to the sample length (easy axis) and perpendicularly to it. The easy axis magnetization, as should be, does not produce any magnetostrictive effect. The maximum laser deflection in the saturating perpendicular field is $\Delta \theta = \unit{0.15}{\ \milli \rad}$  giving the sample curvature $\Delta (1/R)= \Delta \theta/2L=\unit{4.3}{\ \times \millid \reciprocal \meter}$   (distance between the support and the laser spot on the sample is $L=\unit{17.5}{\ \milli \meter}$).
 From these measurements we  obtain   $B^{\gamma,2}=\unit{6.8}{\ \mega \pascal}$ using  eq. (\ref{equ_Delta_antic}) with following parameters: $E_s= \unit{60}{\ \giga \pascal}$, $\nu_s=0.2$, $d= \unit{145}{\ \micro \meter}$, $t=\unit{110}{\ \nano \meter} $ .

This value is in good agreement with the value obtained above ($\unit{6.3}{\ \mega \pascal} $ ) using the inverse technique. It is difficult to get better correspondence of two techniques  taking into account that there are parameters which enter to the calculation for only one of the two techniques:
for the direct technique - elastic moduli of the substrate $E_s$ and $\nu_s$, sample length $L$, other dimensions of the installation and the film thickness $t$;
for the inverse technique - saturation magnetization of the film $M_s$, curvature calibration $1/R$ and determination of $H_K$. Another important difference is that the substrate thickness enters as $d^2$ to the direct technique and as $1/d$ to the inverse one.
Eventual influence of the inhomogeneity on $B^{\gamma,2}$ also can not be excluded: the direct technique provides the average value over the whole sample whereas the inverse technique measures a local value.

\section{Conclusion}

The comparison of direct and inverse ("indirect") techniques of the film magnetostriction measurements shows that both techniques give exactly the same magnetostrictive stress $\sigma^\lambda$ (or $B^{\gamma,2}= -2/3  \sigma^\lambda$) provided all  different experimental conditions are correctly taken into account.

It is natural to characterize the thin films magnetostriction  by  $\sigma^\lambda$ since the lateral deformations of the film are blocked by the substrate and one can measure only the stress (see Fig. \ref{fig_MS1}b). This is in contrast to measurements of the magnetostrictive  deformation $\lambda_s$ of bulk materials (see Fig. \ref{fig_MS1}a).

The key factor to determine the exact expression for $\sigma^\lambda$ to be used is the shape of the curved cantilever realized in the particular conditions of the experiment.
Both the elastic and the magnetoelastic energies depend on the peculiarities of  the curved cantilever shape. Normally a small  cantilever bending involves anticlastic motion of its side edges.

This motion can be hindered by two effects: clamping by the support and non linear geometrical edge blocking for  large cantilever bending. The critical curvature above which  this cylindrical bending with completely flat transversal profile appears is $W^2/Rd \gtrsim 10$  (more precisely anticlastic parameter $\beta  \gtrsim 3$).
For the direct measurements the critical curvature can be achieved only for extremely thin cantilevers  with
 $d\lesssim \sqrt[3]{\lambda_s W^2 t}$.
This critical value is obtained from eq. (\ref{equ_Delta_antic}) neglecting the difference of the elastic moduli of the film and the substrate.  For $\lambda_s=10^{-4}$, $W=\unit{5}{\ \milli \meter}$  and $t=\unit{110}{\ \nano \meter}$,  as an example, this gives $d \lesssim  \unit{6}{\ \micro \meter}$.
The indirect measurements are limited by the maximum strain that can be obtained on the cantilever surface without risk to break it ($e_{max} \sim 2 \times 10^{-3}$). So, in this case the critical limit to the cantilever  thickness is  less severe: $d \lesssim \sqrt{0.2 e_{max}} W$ i.e. $d \lesssim \unit{100}{\ \micro \meter}$ for the same numerical values.

In the extreme case of strongly curved thin cantilever clamped at one end the result of the local measurements of the induced anisotropy will vary not only with the distance from the support but also across the sample width: $K=K(x,y)$ (Appendix \ref{app_shape})

Comparing the  advantages of the two techniques one can think that the indirect technique is more sensitive and easier to realize and practically unlimited by the film thickness. This is true only for the materials with weak magnetic crystalline anisotropy. When this anisotropy is much larger than the maximum induced anisotropy $\lambda_s E_f e_{max} $ this technique loses its advantages.  The magnetostriction measurement of highly anisotropic materials by the direct technique is limited mainly by the availability of strong enough magnetic field ($H>H_K$).
In order to measure materials with low $\lambda_s$ or very thin films by the direct technique one can  reduce the  substrate thickness.

\section*{Acknowledgments}

D. Dekadjevi  is acknowledged for the development of the MOKE experiment.



\appendix
\section*{APPENDICES}
\section{Effect's of the Poisson's ratio}
\label{app_Poisson}

\begin{table}[h]
\begin{tabular}{|p{3cm}|p{4.5cm}|}
\hline
configuration 1 &\\
 \multirow{6}{*}{\includegraphics[angle=0,width=3.0cm]{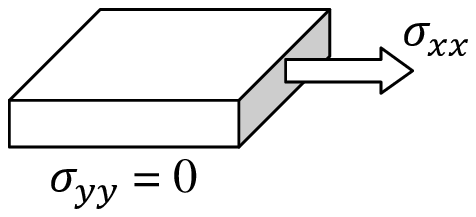}}
& Free edge load:\\
& definition of the\\
&Young's  modulus $E$ \\
& and the Poisson ratio $\nu$ \\
& $e_{xx}=\frac{\sigma_{xx}}{E} $\\
& $e_{yy}=- \nu e_{xx}$\\
    \hline
configuration 2 &\\
 \multirow{4}{*}{\includegraphics[angle=0,width=3.0cm]{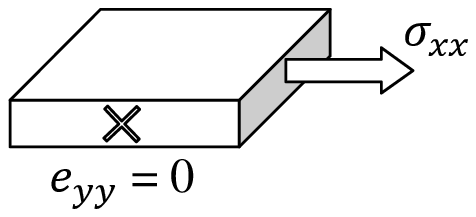}}
& Fixed edge load \\
&$e_{xx}=\frac{1-\nu^2}{E} \sigma_{xx}$ \\
& $\sigma_{yy}=+\nu \sigma_{xx}$\\
&\\
    \hline
configuration 3 &\\
\multirow{4}{*}{\includegraphics[angle=0,width=3.0cm]{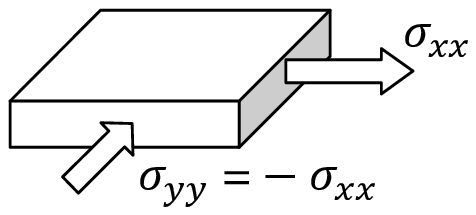}}
& Antisymmetric load/deformation\\
&$e_{xx}=-e_{yy}= \frac{1+\nu}{E} \sigma_{xx}= \frac{\sigma_{xx}}{2G}$\\
&($G$- shear modulus)\\
&\\
    \hline
configuration 4 &\\
 \multirow{4}{*}{\includegraphics[angle=0,width=3.0cm]{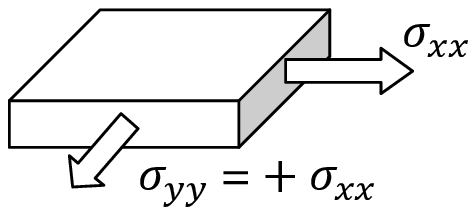}}
& Symmetric load/deformation \\
& $e_{xx}=e_{yy}= \frac{1-\nu}{E} \sigma_{xx}$\\
&(Case of isotropic lateral stresses)\\
&\\
\hline
\end{tabular}
\label{table:nu-summary}
\caption{Summary of characteristic appearances of Poisson's ratio $\nu$ }
\end{table}

Even in the simple case of isotropic materials the appearance of the Poisson's ratio $\nu$   in different formula related to the magneto-elastic interactions presents sometimes a "mystery". In order to simplify this issue we give here a simple summary of relations between stresses $\sigma_{ij}$ and deformations  $e_{ij} $  $(i,j=x,y,z)$ of an isotropic material for typical conditions of mechanical loads useful for the thin film geometry with their proper characteristic appearance of  $\nu$.
Let's consider a small square element with the principal load/deformation along the horizontal $x$ axis  and $z$ axis  perpendicular to the film surface where the normal stress  $\sigma_{zz}  = 0$.

When the side edges of the element are free ($\sigma_{yy}=0$) the lateral linear deformations $e_{xx}$ and  $e_{yy}$ under uniaxial stress $\sigma_{xx}$ define the Young's modulus $E$ and the Poisson's ratio $\nu$ (see configuration 1 in table \ref{table:nu-summary}).  The ratio $e_{yy}/e_{xx}=-\nu$ is directly related to the anticlastic bending of narrow cantilevers with free side edges.

The elastic properties of the element with fixed side edges $e_{yy}=0$ can be obtained  through a linear superposition of two crossed loads ($\sigma_{xx}$ and $\sigma_{yy})$ with free edge conditions presented above, where $\sigma_{yy}$ compensates the transverse deformation produced by principal load $\sigma_{xx}$ : $\sigma_{yy} = \nu \sigma_{xx}$ (configuration 2 in table \ref{table:nu-summary}). The resulting strengthening $\sigma_{xx}/e_{xx}$ of the axial deformation   $(1-\nu^2)^{-1}$ corresponds to the strengthening of strongly bent wide cantilevers where the condition $e_{yy}=0$ is due to the transverse flattening caused by the geometrical edge blocking (see Appendix \ref{app_antic}).

Another useful  superposition of two uniaxial loads $\sigma_{xx}$ and $\sigma_{yy}=-\sigma_{xx}$ (antisymmetric case) is equivalent to the pure  shear deformation along the diagonal of the element (configuration 3 in table \ref{table:nu-summary}). It appears in the magnetoelastic coupling where only the difference between two states with stress axes rotating by $90^\circ$  is essential.

Consequently the elastic parameters of the material  appear in the form $E/(1+\nu)$ in these cases (see eq.(\ref{eq-sigma-lambda}) and (\ref{equ_Delta_antic})).

For completeness  we present also the symmetric superposition of $\sigma_{xx}$ and $\sigma_{yy}=\sigma_{xx}$ which correspond to the thermal and epitaxial stresses of thin films (configuration 4 in table \ref{table:nu-summary}). Factor $(1-\nu)$ obtained here instead of the factor $(1+\nu)$ in the previous antisymmetric configuration  enter to the expression of the substrate curvature (eq. (\ref{equ_Delta_antic})) for  this case well known as Stoney's formula \cite{Janssen-09}.

\section{Evolution of anticlastic deformation}
\label{app_antic}

\begin{figure}[h]
\includegraphics[angle=0,width=7.0cm]{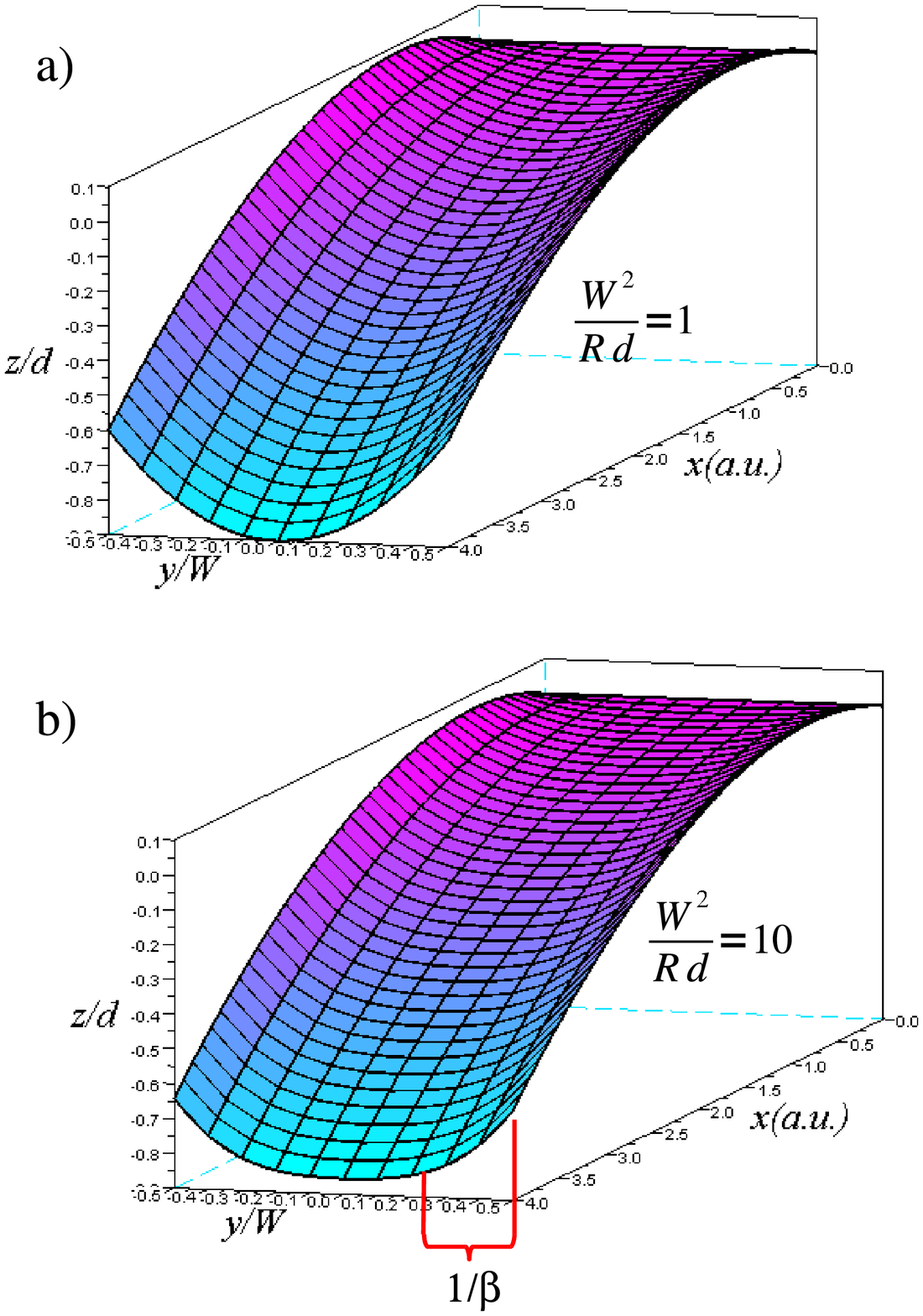}
\caption {(Color on line) Anticlastic deformation of a bent cantilever for different values of the reduced curvature $W^2/R d$.
a) small enough bending $W^2/R d =1$ (anticlastic parameter $\beta = 1.28$)
b) strong bending $W^2/R d =10$ ($\beta = 4$) with a flattening in the center.
Transverse profile calculated from eq. (\ref{equ_z}). (z/d) is differently magnified in a) and b) for clearness. We added to this figure  a gradual flattening qualitatively reproducing the clamping effect of the support (at $x=0$).}
\label{fig-shape}
\end{figure}

The details of induced  anisotropy of  magnetostrictive  films  on curved substrates related to their anticlastic deformation, clamping effect were largely   discussed in the literature \cite{lacheisserie-jmmm94,Marcus_JMMM_97,Iannotti-JMMM99,Dean-06}. All these numerical and analytical solutions were limited to very small deformations. It is interesting to consider the effect of the evolution of the form of a cantilever with increasing curvature.
The anticlastic deformation can not be maintained for strong curvatures because of appearance of tear stresses at the side edges of the cantilever. General equations describing thin plate 2D profile  $Z(x,y)$  are  non-linear. For our cantilever with principal curvature $1/R$ along its length ($Z(x,y)=-x^2/2R + z(y)$) there is an analytical  solution: the transverse profile of the vertical deviation $z(y)$ can be obtained from the 4th order  differential equation \cite{Ashwell-50,Conway-65}:

\begin{eqnarray}
\frac{d^4 z}{dy^4} + 4 \left( \frac{\beta}{W}  \right)^4    z = 0
\end{eqnarray}

where
\begin{eqnarray}
\beta^2= \frac{W^2}{R \ d} \left[3(1-\nu^2)\right]^{1/2}
\end{eqnarray}
is a parameter (anticlastic parameter) which determines the degree of the non-linearity .\\
$W$ is the cantilever width  and $d$ its thickness.

The solution of this equation is :

\begin{eqnarray}
\nonumber
\frac{z}{d}&=\frac{\nu}{\sqrt{3(1-\nu^2)}} \left\{
C_{-} \cosh\left( \beta \frac{y}{W} \right) \cos \left(  \beta \frac{y}{W}  \right) \right. \\
&\left. + C_{+}\sinh\left( \beta \frac{y}{W} \right) \sin \left( \beta \frac{y}{W} \right)
\right\}
\label{equ_z}
\end{eqnarray}

with

$$C_{\substack{-\\+}}=\frac{ \sinh(\beta/2) \cos(\beta/2) \mp \cosh(\beta/2) \sin(\beta/2)}{\sinh(\beta)+\sin(\beta)}$$

For  $\beta\ll 1$, it gives the anticlastic transversal curvature ($\nu/R)$  opposite  to the principal curvature of the cantilever ($-1/R$):
\begin{eqnarray}
z(y)=\frac{\nu}{2R}(y^2-W^2/8)
\label{eq_parabol}
\end{eqnarray}
as illustrated in Fig. \ref{fig-shape}. It is interesting to note that this parabolic transversal deformation proportional to $1/R$  persists till $\beta \sim 1$.

\begin{figure}[h]
\includegraphics[angle=0,width=6.5cm]{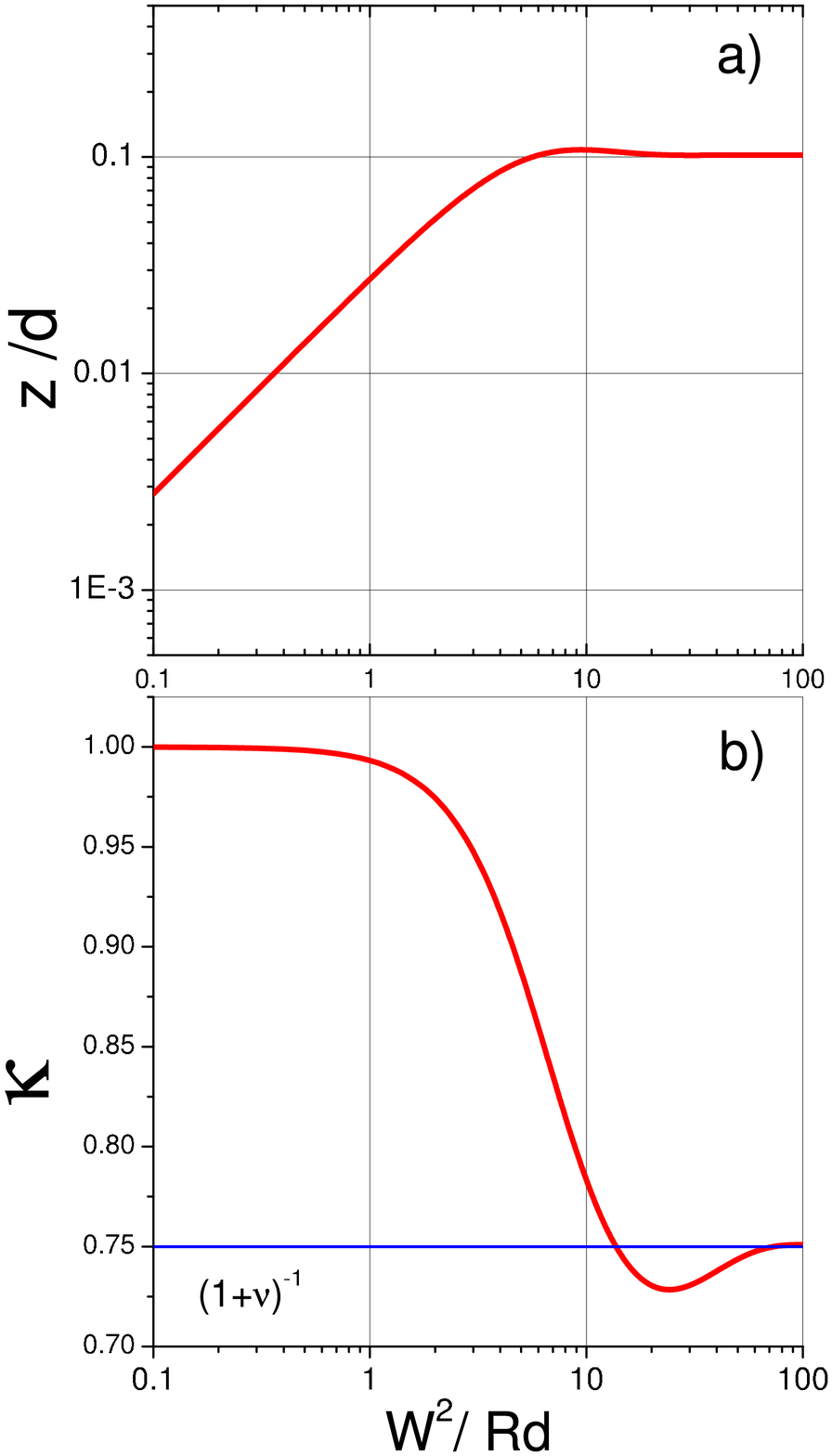}
\caption{(Color on line)
Effects of the reduced curvature $ \nicefrac{W^2}{R d}$:
 a) Transverse deviation $z$ at the edge of the cantilever ($y=W/2$) relatively to its thickness $d$. b) Correction for induced anisotropy  in the center of a wide cantilever compared to an equally bended narrow cantilever.  $\nu=1/3$ is used for the calculations.}
\label{fig-z-kappa}
\end{figure}

For larger $\beta$ the cantilever starts to flatten in the center.  For $\beta \gtrsim 3$ (see Fig. \ref{fig-shape} ) the anticlastic deformation remains only at the sample edges in a region of $W/\beta$ width and its amplitude  saturates (Fig. \ref{fig-z-kappa}-a). This edge deformation region is clearly demonstrated on images  of bent metallic sheets shown by Conway and Nickola  \cite{Conway-65}.
The transition from anticlastic to cylindrical bending produces corresponding modification of the ratio  between induced anisotropy and the principal cantilever curvature $1/R$
[See changes of the forms of the effective coefficient of magnetostrictive coupling $B$ in eq. (\ref{equ_Delta_antic}) and (\ref{equ_Delta_cyl})].

It is interesting to follow the whole evolution of this ratio from small (linear regime)   to large (non linear regime) cantilever bending.

The induced anisotropy is determined by $(e_{xx}-e_{yy})$ at the cantilever surface where the film is deposited:
\begin{eqnarray}
e_{xx}&=&\frac{d}{2} \frac{\partial^2 Z}{\partial x^2}= -\frac{d}{2} \frac{1}{R}\\
e_{yy}&=& \frac{d}{2} \frac{\partial^2 Z}{\partial y^2} = \frac{d}{2} \frac{d^2 z}{d y^2}
\end{eqnarray}
For a narrow cantilever  $e_{yy}=-\nu e_{xx}$ is valid for any curvature ($\beta \ll 1$  see eq. (\ref{eq_parabol})).

The induced anisotropy measured in the center of  wide cantilever of the same curvature must be corrected by a factor:

\begin{eqnarray}
\kappa(1/R)=  \frac{e_{xx}-e_{yy}}{(1+\nu)e_{xx}} =\frac{1}{1+\nu} \left( 1+ R  \left. \frac{d^2 z}{d y^2}\right|_{y=0} \right)
\label{equ-kappa}
\end{eqnarray}

For $\nu=1/3$ the evolution of $\kappa$ is shown in Fig. \ref{fig-z-kappa}- b).

When bending is small ($1/R \lesssim W/d^2$ or $\beta \lesssim 1$ ) the anticlastic shape does not depend on the cantilever width $d^2 z/dy^2=\nu /R$ and $\kappa=1$. When bending is large ($1/R\gtrsim 10 W/d^2$ or $\beta \gtrsim 3 $) and the  center of the cantilever becomes flat ($d^2 z/dy^2 \approx 0$) we have to correct the usual expression of the induced anisotropy by  $\kappa = 1/(1+\nu)$.

\section{Stress variation by the cantilever thickness and length}
\label{app_shape}

There is a large  variety of the stress distribution for different cases of the cantilever deformation.
We consider here two specific cases relevant to our experiment. In this appendix, for simplicity, we discuss only the free edge linear curvature of the cantilever corresponding to configuration 1 in Appendix \ref{app_Poisson} with $\sigma_{xx}=E_s e_{xx}$. Here we have to distinguish Young modulus and Poisson's ratio for the substrate ($E_s$ and $\nu_s$) and for the magnetostrictive film ($E_f$ and $\nu_f$).
The latter appear only in $B^{\gamma,2}= -3/2~ \lambda_s ~ E_f/(1+\nu_f)$.

\begin{figure}[h]
\includegraphics[angle=0,width=9.0cm]{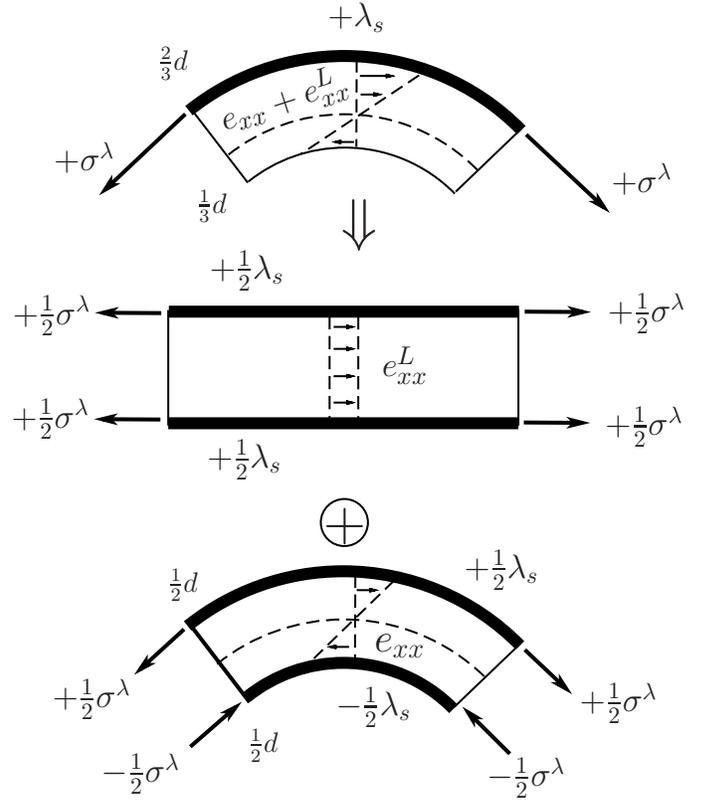}
\caption{
The practical experimental configuration  with a film deposited on one substrate surface can be considered as the superposition of a symmetric problem with 2 equivalent films deposited on both surfaces (sample lengthening only) and an antisymmetric problem with 2 films of opposite magnetostriction on both surfaces (sample curvature only).}
\label{fig-sym-asym}
\end{figure}

The deformation of the cantilever under the magnetostrictive  stress of the film with magnetostriction $+\lambda_s$  deposited onto only one surface is not symmetrical with respect to its midplane. The corresponding effect of the shift of the neutral line  of the elastic plate deformation somewhat complicates the analysis provided in different papers \cite{lacheisserie-jmmm94,Marcus_JMMM_97}.
This practical problem  can be considered as  the linear superposition of  2 simpler problems: symmetric - the  deformation of  the substrate by   2 equivalent films deposited on both surfaces with magnetostriction coefficient $+1/2~\lambda_s$, and antisymmetric - 2 films with opposite magnetostrictions  coefficients $+1/2~\lambda_s$ and $-1/2~\lambda_s$ on opposite cantilever surfaces
(see Fig. \ref{fig-sym-asym} ).

Let's consider  the influence of the longitudinal stress $\sigma_{xx}^\parallel~=~ \sigma^\lambda$ of the film magnetized along the $x$ axis as indicated in Fig. \ref{fig-sym-asym}.

In the symmetric case, evidently, the only deformation is lengthening $e_{xx}^L$ and there is no curvature. From the balance of the forces between the constant magnetostrictive stress on both surfaces and the Hooke's force of the sample lengthening one gets:
\begin{eqnarray}
t \ \sigma^\lambda = d\ E_s \ e_{xx}^L
\label{equ_Hooke}
\end{eqnarray}

In the antisymmetric case the pure curvature ($1/R$) with the neutral line at the middle of cantilever thickness is produced without any lengthening of the sample, i.e. $e_{xx}(z)=z/R$. Here one should balance  the momentum of surface magnetostrictive forces and the momentum of the bulk Hooke's forces relative to the neutral central line:
\begin{eqnarray}
t \ \sigma^\lambda \ d/2 = \int_{-d/2}^{d/2}  z \ E_s e_{xx}(z) dz
\label{equ_momenta}
\end{eqnarray}

that gives the resulting curvature:
\begin{eqnarray}
\frac{1}{R(\sigma_{xx}^\parallel)}=\frac{6 t \sigma^\lambda }{E_s d^2}
\end{eqnarray}

For the initial combined problem, the deformation of curvature and lengthening have the same origin and are related by the following relation:
\begin{eqnarray}
e_{xx}(z)=6 \left( \frac{z}{d} \right)e_{xx}^L
\label{equ_L}
\end{eqnarray}
Thus the maximal value of $e_{xx}(z)$ is $3 |e_{xx}^L |$  and the neutral line of the total deformation $e_{xx}(z) + e_{xx}^L$ is located at $z=-d/6$ (See Fig. \ref{fig-sym-asym}).

For complete description  of the effect of the magnetization rotation more components of the stress should be considered: the tranversal stress for the magnetization along $x$ axis ${\sigma_{yy}^\parallel~=~-1/2~\sigma^\lambda}$ and the stresses for the perpendicular magnetization ${\sigma_{xx}^\perp~=~-~1/2~\sigma^\lambda}$ and  $\sigma_{yy}^\perp ~ =~  \sigma^\lambda$. The position of the neutral plane for each component is the same and their total influence increases curvature $1/R(\sigma_{xx}^\parallel)$ by $3(1+\nu_s)/2$ found in eq. (\ref{equ_Delta_antic}).
 The shift of the neutral plane  is found not only for magnetostrictive film stress \cite{lacheisserie-jmmm94,Marcus_JMMM_97,Iannotti-JMMM99,Sander-RPP99} but also for thermal or epitaxial stresses (\cite{Janssen-09}  and references therein).

Second question we have to address is the stress distribution and the corresponding curvature variation along the cantilever length.
The simplest way to obtain this is to balance the momentum of forces around a given point of the cantilever curved by force $F$ applied to its free end ($x=L$), the other end ($x=0$) being clamped.

Let's consider the equilibrium of  part $[x,L]$ of the cantilever around the middle point at its left edge.
The momentum of  bulk Hooke's forces (see the second term of eq. (\ref{equ_momenta}) ) applied to $x$ should be equal to the applied force momentum.
\begin{eqnarray}
 \int_{-d/2}^{d/2}  z \ E_s e_{xx}(z) dz = \frac{F}{W}(L-x)
\end{eqnarray}
that gives:
\begin{eqnarray}
\frac{1}{R}= \frac{12 F}{E_S W d^3}  (L-x)
\end{eqnarray}

This simple solution is valid when the sample side edges are free and the anticlastic deformation occurs everywhere along the sample  length.
In reality the anticlastic deformation is blocked at the sample support (clamping) with corresponding bending  hardening by a factor $1-\nu_s^2$ ( see configuration 2 in Appendix \ref{app_Poisson} and modification of  elastic coefficient $A$ in eq. (\ref{equ_nrj}) from anticlastic to cylindrical bending). This phenomena can be described qualitatively by introducing an effective Poisson's ratio correction $\tilde{\nu}(x)= \nu_s \exp(-x/a)$ where $a \sim W/2 $ is the characteristic relaxation length   of the clamping effect.
With this correction:
\begin{eqnarray}
\frac{1}{R}= \frac{12 F}{E_S W d^3} (1-\tilde{\nu}^2(x)) (L-x)
\end{eqnarray}

The clamping produces  even stronger influence onto the induced magnetic anisotropy that has to be multiplied  by $(1+\tilde\nu)$ according to modification of  the effective magnetoelastic  coefficient $B$ in  eq. (\ref{equ_nrj}) from anticlastic to cylindrical bending.
So we obtain:
\begin{eqnarray}
B(x)=\frac{1}{2} d B^{\gamma,2} \frac{1+\nu_s}{1+ \tilde{\nu}(x)}
\end{eqnarray}
(see eq. (\ref{equ_Wms})).

Combining $1/R(x)$ and $B(x)$ in eq. (\ref{equ_Keff}) we get a non-linear variation  of the anisotropy with the distance from the cantilever support:

\begin{eqnarray}
\nonumber
K_{eff}(x)= K+\\
 \frac{6 F  B^{\gamma,2} }{E_S W d^2} (1+\nu_s)(1-\tilde{\nu}(x)) (L-x)
\label{equ_Knorm}
\end{eqnarray}

instead of the traditionally considered linear variation without clamping effect ($\tilde{\nu}~=~0$).



\bibliographystyle{apsrev}

\end{document}